\documentclass[aps,preprint,nofootinbib,superscriptaddress]{revtex4}%
\usepackage{amssymb}
\usepackage{amsmath}
\usepackage{epsfig}
\usepackage{amsfonts}
\usepackage{graphicx}
\usepackage{hyperref}%
\providecommand{\U}[1]{\protect\rule{.1in}{.1in}}
\begin{document}
\title{Biquaternion Construction of SL(2,C) Yang-Mills Instantons}
\author{Jen-Chi Lee}
\email{jcclee@cc.nctu.edu.tw}
\affiliation{Department of Electrophysics, National Chiao-Tung University, Hsinchu, Taiwan, R.O.C.}
\date{\today }

\begin{abstract}
We use biquaternion to construct SL(2,C) ADHM Yang-Mills instantons. The
solutions contain 16k-6 moduli parameters for the kth homotopy class, and
include as a subset the SL(2,C) (M,N) instanton solutions constructed
previously. In constrast to the SU(2) instantons, the SL(2,C) instantons
inhereit jumping lines or singulariries which are not gauge artifacts and can
not be gauged away.

\end{abstract}
\preprint{ }
\maketitle
\tableofcontents

%

\setcounter{equation}{0}
\renewcommand{\theequation}{\arabic{section}.\arabic{equation}}%

\section{Introduction}

The classical exact solutions of Euclidean $SU(2)$ (anti)self-dual Yang-Mills
(SDYM) equation were intensively studied by pure mathematicians and
theoretical physicists in 1970s. The first BPST $1$-instanton solution
\cite{BPST} with $5$ moduli parameters was found in 1975. The CFTW k-instanton
solutions \cite{CFTW} with $5k$ moduli parameters were soon constructed, and
then the number of moduli parameters of the solutions for each homotopy class
$k$ was extended to $5k+4$ ($5$,$13$ for $k=1$,$2$) \cite{JR} based on the
conformal symmetry of massless pure YM equation. The complete solutions with
$8k-3$ moduli parameters for each $k$-th homotopy class were finally worked
out in 1978 by mathematicians ADHM \cite{ADHM} using theory in algebraic
geometry. Through an one to one correspondence between anti-self-dual
SU(2)-connections on $S^{4}$ and holomorphic vector bundles on $CP^{3}$, ADHM
converted the highly nontrivial anti-SDYM equations into a much more simpler
system of quadratic algebraic equations in quaternions. The explicit closed
form of the complete solutions for $k=2,3$ had been worked out \cite{CSW}.

There are many important applications of instantons to algebraic geometry and
quantum field theory. One important application of instantons in algebraic
geometry was the classification of four-manifolds \cite{5}. On the physics
side, the non-perturbative instanton effect in QCD resolved the $U(1)_{A}$
problem \cite{U(1)}. Another important application of YM instantons in quantum
field theory was the introduction of $\theta$- vacua \cite{the} in
nonperturbative QCD, which created the strong $CP$ problem.

In addition to $SU(2)$, the ADHM construction has been generalized to the
cases of $SU(N)$ SDYM and many other SDYM theories with compact Lie groups
\cite{CSW,JR2}. In this talk we are going to consider the classical solutions
of non-compact $SL(2,C)$ SDYM system. YM theory based on $SL(2,C)$ was first
discussed in 1970s \cite{WY,Hsu}. It was found that the complex $SU(2)$ YM
field configurations can be interpreted as the real field configurations in
$SL(2,C)$ YM theory. However, due to the non-compactness of $SL(2,C)$, the
Cartan-Killing form or group metric of $SL(2,C)$ is not positive definite.
Thus the action integral and the Hamiltonian of non-compact $SL(2,C)$ YM
theory may not be positve. Nevertheless, there are still important motivations
to study $SL(2,C)$ SDYM theory. For example, it was shown that the $4D$
$SL(2,C)$ SDYM equation can be dimensionally reduced to many important $1+1$
dimensional integrable systems \cite{Mason}, such as the KdV equation and the
nonlinear Schrodinger equation.

\section{SL(2,C) SDYM Equation}

We first briefly review the $SL(2,C)$ YM theory. It was shown that \cite{WY}
there are two linearly independent choices of $SL(2,C)$ group metric\newline%

\begin{equation}
g^{a}=%
\begin{pmatrix}
I & 0\\
0 & -I
\end{pmatrix}
,g^{b}=%
\begin{pmatrix}
0 & I\\
I & 0
\end{pmatrix}
\end{equation}
where $I$ is the $3\times3$ unit matrix. In general, we can choose
\begin{equation}
g=\cos\theta g^{a}+\sin\theta g^{b}%
\end{equation}
where $\theta$ = real constant. Note that the metric is not positive definite
due to the non-compactness of $SL(2,C).$ On the other hand, it was shown that
$SL(2,C)$ group can be decomposed such that \cite{Lee}%
\begin{equation}
SL(2,C)=SU(2)\cdot P,P\in H\newline%
\end{equation}
\newline where $SU(2)$ is the maximal compact subgroup of $SL(2,C)$, $P\in H$
(not a group) and $H=\{P|P$ is Hermitain, positive definite, and $detP=1\}$.
The parameter space of $H$ is a noncompact space $R^{3}$. The third homotopy
group is thus \cite{Lee}%
\begin{equation}
\pi_{3}[SL(2,C)]=\pi_{3}[S^{3}\times R^{3}]=\pi_{3}(S^{3})\cdot\pi_{3}%
(R^{3})=Z\cdot I=Z\newline\newline%
\end{equation}
\newline where $I$ is the identity group, and $Z$ is the integer
group.\newline\qquad On the other hand, Wu and Yang \cite{WY} have shown that
a complex $SU(2)$ gauge field is related to a real $SL(2,C)$ gauge field.
Starting from $SU(2)$ complex gauge field formalism, we can write down all the
$SL(2,C)$ field equations. Let
\begin{equation}
G_{\mu}^{a}=A_{\mu}^{a}+iB_{\mu}^{a}%
\end{equation}
and, for convenience, we set the coupling constant $g=1$. The complex field
strength is defined as
\begin{equation}
F_{\mu\nu}^{a}\equiv H_{\mu\nu}^{a}+iM_{\mu\nu}^{a},a,b,c=1,2,3
\end{equation}
where
\begin{align}
H_{\mu\nu}^{a}  &  =\partial_{\mu}A_{\nu}^{a}-\partial_{\nu}A_{\mu}%
^{a}+\epsilon^{abc}(A_{\mu}^{b}A_{\nu}^{c}-B_{\mu}^{b}B_{\nu}^{c}),\nonumber\\
M_{\mu\nu}^{a}  &  =\partial_{\mu}B_{\nu}^{a}-\partial_{\nu}B_{\mu}%
^{a}+\epsilon^{abc}(A_{\mu}^{b}B_{\nu}^{c}-A_{\mu}^{b}B_{\nu}^{c}),
\end{align}
then $SL(2,C)$ Yang-Mills equation can be written as
\begin{align}
\partial_{\mu}H_{\mu\nu}^{a}+\epsilon^{abc}(A_{\mu}^{b}H_{\mu\nu}^{c}-B_{\mu
}^{b}M_{\mu\nu}^{c})  &  =0,\nonumber\\
\partial_{\mu}M_{\mu\nu}^{a}+\epsilon^{abc}(A_{\mu}^{b}M_{\mu\nu}^{c}-B_{\mu
}^{b}H_{\mu\nu}^{c})  &  =0.
\end{align}
The $SL(2,C)$ SDYM equations are%
\begin{align}
H_{\mu\nu}^{a}  &  =\frac{1}{2}\epsilon_{\mu\nu\alpha\beta}H_{\alpha\beta
},\nonumber\\
M_{\mu\nu}^{a}  &  =\frac{1}{2}\epsilon_{\mu\nu\alpha\beta}M_{\alpha\beta}.
\label{self}%
\end{align}
The Yang-Mills Equation above can be derived from the following Lagrangian%
\begin{equation}
L_{\theta}=\frac{1}{4}[F_{\mu\nu}^{i}]^{T}g_{ij}[F_{\mu\nu}^{j}]=\cos{\theta
}(\frac{1}{4}H_{\mu\nu}^{a}H_{\mu\nu}^{a}-\frac{1}{4}M_{\mu\nu}^{a}M_{\mu\nu
}^{a})+\sin{\theta}(\frac{1}{2}H_{\mu\nu}^{a}M_{\mu\nu}^{a}) \label{action}%
\end{equation}
where $F_{\mu\nu}^{k}=H_{\mu\nu}^{k}$ and $F_{\mu\nu}^{3+k}=M_{\mu\nu}^{k}$
for $k=1,2,3$. Note that $L_{\theta}$ is indefinite for any real value
$\theta$. We shall only consider the particular case for $\theta=0$ in this
talk, i.e.
\begin{equation}
L=\frac{1}{4}(H_{\mu\nu}^{a}H_{\mu\nu}^{a}-M_{\mu\nu}^{a}M_{\mu\nu}^{a}),
\end{equation}
for the action density in discussing the homotopic classifications of our solutions.

\section{Biquaternion construction of $SL(2,C)$ YM Instantons}

Instead of quaternion in the $Sp(1)$ ($=SU(2)$) ADHM construction, we will use
\textit{biquaternion} to construct $SL(2,C)$ SDYM instantons. A quaternion $x$
can be written as%
\begin{equation}
x=x_{\mu}e_{\mu}\text{, \ }x_{\mu}\in R\text{, \ }e_{0}=1,e_{1}=i,e_{2}%
=j,e_{3}=k \label{x}%
\end{equation}
where $e_{1},e_{2}$ and $e_{3}$ anticommute and obey%
\begin{align}
e_{i}\cdot e_{j}  &  =-e_{j}\cdot e_{i}=\epsilon_{ijk}e_{k};\text{
\ }i,j,k=1,2,3,\\
e_{1}^{2}  &  =-1,e_{2}^{2}=-1,e_{3}^{2}=-1.
\end{align}

A (ordinary) biquaternion (or complex-quaternion) $z$ can be written as%
\begin{equation}
z=z_{\mu}e_{\mu}\text{, \ }z_{\mu}\in C,
\end{equation}
which will be used in this talk. Occasionally $z$ can be written as%
\begin{equation}
z=x+yi
\end{equation}
where $x$ and $y$ are quaternions and $i=\sqrt{-1},$ not to be confused with
$e_{1}$ in Eq.(\ref{x}). For biquaternion, the biconjugation \cite{Ham}%
\begin{equation}
z^{\circledast}=z_{\mu}e_{\mu}^{\dagger}=z_{0}e_{0}-z_{1}e_{1}-z_{2}%
e_{2}-z_{3}e_{3}=x^{\dagger}+y^{\dagger}i,
\end{equation}
will be heavily used in this talk. In contrast to the real number norm square
of a quaternion, the norm square of a biquarternion used in this talk is
defined to be%
\begin{equation}
|z|_{c}^{2}=z^{\circledast}z=(z_{0})^{2}+(z_{1})^{2}+(z_{2})^{2}+(z_{3})^{2}%
\end{equation}
which is a \textit{complex} number in general as a subscript $c$ is used in
the norm.

We are now ready to proceed the construction of $SL(2,C)$ instantons. We begin
by introducing the $(k+1)\times k$ biquarternion matrix $\Delta(x)=a+bx$%

\begin{equation}
\Delta(x)_{ab}=a_{ab}+b_{ab}x,\text{ }a_{ab}=a_{ab}^{\mu}e_{\mu},b_{ab}%
=b_{ab}^{\mu}e_{\mu} \label{ab}%
\end{equation}
where $a_{ab}^{\mu}$ and $b_{ab}^{\mu}$ are complex numbers, and $a_{ab}$ and
$b_{ab}$ are biquarternions. The biconjugation of the $\Delta(x)$ matrix is
defined to be%
\begin{equation}
\Delta(x)_{ab}^{\circledast}=\Delta(x)_{ba}^{\mu}e_{\mu}^{\dagger}%
=\Delta(x)_{ba}^{0}e_{0}-\Delta(x)_{ba}^{1}e_{1}-\Delta(x)_{ba}^{2}%
e_{2}-\Delta(x)_{ba}^{3}e_{3}.
\end{equation}

In contrast to the of $SU(2)$ instantons, the quadratic condition of $SL(2,C)$
instantons reads%

\begin{equation}
\Delta(x)^{\circledast}\Delta(x)=f^{-1}=\text{symmetric, non-singular }k\times
k\text{ matrix for }x\notin J\text{,} \label{ff}%
\end{equation}
from which\ we can deduce that $a^{\circledast}a,b^{\circledast}%
a,a^{\circledast}b$ and $b^{\circledast}b$ are all symmetric matrices. We
stress here that it will turn out the choice of \textit{biconjugation}
operation is crucial for the follow-up discussion in this work. On the other
hand, for $x\in J,$ $\det\Delta(x)^{\circledast}\Delta(x)=0$. The set $J$ is
called singular locus or "jumping lines" in the mathematical literatures and
was discussed in \cite{LLT}. In contrast to the $SL(2,C)$ instantons, there
are no jumping lines for the case of $SU(2)$ instantons. In the $Sp(1)$
quaternion case, the symmetric condition on $f^{-1}$ means $f^{-1}$ is real.
For the $SL(2,C)$ biquaternion case, however, it can be shown that symmetric
condition on $f^{-1}$ implies $f^{-1}$ is \textit{complex}.

To construct the self-dual gauge field, we introduce a $(k+1)\times1$
dimensional biquaternion vector $v(x)$ satisfying the following two conditions%
\begin{align}
v^{\circledast}(x)\Delta(x)  &  =0,\label{null}\\
v^{\circledast}(x)v(x)  &  =1. \label{norm2}%
\end{align}
Note that $v(x)$ is fixed up to a $SL(2,C)$ gauge transformation%
\begin{equation}
v(x)\longrightarrow v(x)g(x),\text{ \ \ }g(x)\in\text{ }1\times1\text{
Biquaternion}.
\end{equation}
Note also that in general a $SL(2,C)$ matrix can be written in terms of a
$1\times1$ biquaternion as%
\begin{equation}
g=\frac{q_{\mu}e_{\mu}}{\sqrt{q^{\circledast}q}}=\frac{q_{\mu}e_{\mu}}%
{|q|_{c}}.
\end{equation}
The next step is to define the gauge field%

\begin{equation}
G_{\mu}(x)=v^{\circledast}(x)\partial_{\mu}v(x), \label{A}%
\end{equation}
which is a $1\times1$ biquaternion. Note that, unlike the case for $Sp(1)$,
$G_{\mu}(x)$ needs not to be anti-Hermitian.

We can now define the $SL(2,C)$ field strength
\begin{equation}
F_{\mu\nu}=\partial_{\mu}G_{\nu}(x)+G_{\mu}(x)G_{\nu}(x)-[\mu
\longleftrightarrow\nu].
\end{equation}
To show that $F_{\mu\nu}$ is self-dual, one first show that the operator
\begin{equation}
P=1-v(x)v^{\circledast}(x)
\end{equation}
is a projection operator $P^{2}=P$, and can be written in terms of $\Delta$ as%

\begin{equation}
P=\Delta(x)f\Delta^{\circledast}(x). \label{P}%
\end{equation}
The self-duality of $F_{\mu\nu}$ can now be proved as following

$\bigskip$%
\begin{align}
F_{\mu\nu}  &  =\partial_{\mu}(v^{\circledast}(x)\partial_{\nu}%
v(x))+v^{\circledast}(x)\partial_{\mu}v(x)v^{\circledast}(x)\partial_{\nu
}v(x)-[\mu\longleftrightarrow\nu]\nonumber\\
&  =v^{\circledast}(x)b(e_{\mu}e_{\nu}^{\dagger}-e_{\nu}e_{\mu}^{\dagger
})fb^{\circledast}v(x) \label{F}%
\end{align}
where we have used Eqs.(\ref{ab}),(\ref{null}) and (\ref{P}). Finally the
factor $(e_{\mu}e_{\nu}^{\dagger}-e_{\nu}e_{\mu}^{\dagger})$ above can be
shown to be self-dual%
\begin{align}
\sigma_{\mu\nu}  &  \equiv\frac{1}{4i}(e_{\mu}e_{\nu}^{\dagger}-e_{\nu}e_{\mu
}^{\dagger})=\frac{1}{2}\epsilon_{\mu\nu\alpha\beta}\sigma_{\alpha\beta
},\label{duall}\\
\overset{\_}{\sigma}_{\mu\nu}  &  =\frac{1}{4i}(e_{\mu}^{\dagger}e_{\nu
}-e_{\nu}^{\dagger}e_{\mu})=-\frac{1}{2}\epsilon_{\mu\nu\alpha\beta
}\overset{\_}{\sigma}_{\alpha\beta}.
\end{align}
This proves the self-duality of $F_{\mu\nu}.$ We thus have constructed many
$SL(2,C)$ SDYM field configurations.

To count the number of moduli parameters for the $SL(2,C)$ $k$-instantons we
have constructed , one uses transformations which preserve conditions
Eq.(\ref{ff}), Eq.(\ref{null}) and Eq.(\ref{norm2}), and the definition of
$G_{\mu}$ in Eq.(\ref{A}) to bring $b$ and $a$ in Eq.(\ref{ab}) into a simple
canonical form%

\begin{equation}
b=%
\begin{bmatrix}
0_{1\times k}\\
I_{k\times k}%
\end{bmatrix}
, \label{b}%
\end{equation}

\begin{equation}
a=%
\begin{bmatrix}
\lambda_{1\times k}\\
-y_{k\times k}%
\end{bmatrix}
\label{a}%
\end{equation}
where $\lambda$ and $y$ are biquaternion matrices with orders $1\times k$ and
\ $k\times k$ respectively, and $y$ is symmetric%

\begin{equation}
y=y^{T}.
\end{equation}
The constraints for the moduli parameters are%
\begin{equation}
a_{ci}^{\circledast}a_{cj}=0,i\neq j,\text{ and \ }y_{ij}=y_{ji}. \label{dof}%
\end{equation}
The total number of moduli parameters for $k$-instanton can be calculated
through Eq.(\ref{dof}) to be%
\begin{equation}
\text{\# of moduli for }SL(2,C)\text{ }k\text{-instantons}=16k-6,
\end{equation}
which is twice of that of the case of $Sp(1).$ Roughly speaking, there are
$8k$ parameters for instanton "biquaternion positions" and $8k$ parameters for
instanton "sizes". Finally one has to subtract an overall $SL(2,C)$ gauge
group degree of freesom $6.$ This picture will become more clear when we give
examples of explicit constructions of $SL(2,C)$ instantons in the next section.

\section{Examples of $SL(2,C)$ instantons and Jumping lines}

In this section, we will explicitly construct examples of $SL(2,C)$ YM
instantons to illustrate our prescription given in the last section. Example
of $SL(2,C)$ instantons with jumping lines will also be given.

\subsection{The $SL(2,C)$ $(M,N)$ Instantons}

In this first example, we will reproduce from the ADHM construction the
$SL(2,C)$ $(M,N)$ instanton solutions constructed in \cite{Lee}. We choose the
biquaternion $\lambda_{j}$ in Eq.(\ref{a}) to be $\lambda_{j}e_{0}$ with
$\lambda_{j}$ a \textit{complex} number, and choose $y_{ij}=y_{j}\delta_{ij}$
to be a diagonal matrix with $y_{j}=y_{j\mu}e_{\mu}$ a quaternion. That is%

\begin{equation}
\Delta(x)=%
\begin{bmatrix}
\lambda_{1} & \lambda_{2} & ... & \lambda_{k}\\
x-y_{1} & 0 & ... & 0\\
0 & x-y_{2} & ... & 0\\
. & ... & ... & ...\\
0 & 0 & ... & x-y_{k}%
\end{bmatrix}
, \label{delta}%
\end{equation}
which satisfies the constraint in Eq.(\ref{dof}). One can calculate the gauge
potential as%
\begin{align}
G_{\mu}  &  =v^{\circledast}\partial_{\mu}v=\frac{1}{4}[e_{\mu}^{\dagger
}e_{\nu}-e_{\nu}^{\dagger}e_{\mu}]\partial_{\nu}\ln(1+\frac{\lambda_{1}^{2}%
}{|x-y_{1}|^{2}}+...+\frac{\lambda_{k}^{2}}{|x-y_{k}|^{2}})\nonumber\\
&  =\frac{1}{4}[e_{\mu}^{\dagger}e_{\nu}-e_{\nu}^{\dagger}e_{\mu}%
]\partial_{\nu}\ln(\phi)
\end{align}
where
\begin{equation}
\phi=1+\frac{\lambda_{1}^{2}}{|x-y_{1}|^{2}}+...+\frac{\lambda_{k}^{2}%
}{|x-y_{k}|^{2}}. \label{fai}%
\end{equation}
For the case of $Sp(1),$ $\lambda_{j}$ is a real number and $\lambda
_{j}\lambda_{j}^{\dagger}=\lambda_{j}^{2}$ is a real number. So $\phi$ in
Eq.(\ref{fai}) is a complex-valued function in general. If we choose $k=1$ and
define $\lambda_{1}^{2}=\frac{\alpha_{1}^{2}}{1+i},$ then%
\begin{equation}
\phi=1+\frac{\frac{\alpha_{1}^{2}}{1+i}}{|x-y_{1}|^{2}}.
\end{equation}
The gauge potential is%
\begin{align}
G_{\mu}  &  =\frac{1}{4}[e_{\mu}^{\dagger}e_{\nu}-e_{\nu}^{\dagger}e_{\mu
}]\partial_{\nu}\ln(1+\frac{\frac{\alpha_{1}^{2}}{1+i}}{|x-y_{1}|^{2}}%
)=\frac{1}{4}[e_{\mu}^{\dagger}e_{\nu}-e_{\nu}^{\dagger}e_{\mu}]\partial_{\nu
}\ln(1+\frac{\alpha_{1}^{2}}{|x-y_{1}|^{2}}+i)\nonumber\\
&  =\frac{1}{2}[e_{\mu}^{\dagger}e_{\nu}-e_{\nu}^{\dagger}e_{\mu}%
]\frac{-\alpha_{1}^{2}(x-y_{1})_{\nu}}{|x-y_{1}|^{4}+(|x-y_{1}|^{2}+\alpha
_{1}^{2})^{2}}[\frac{|x-y_{1}|^{2}+\alpha_{1}^{2}}{|x-y_{1}|^{2}}-i]
\end{align}
which reproduces the $SL(2,C)$ $(M,N)=(1,0)$ solution calculated in
\cite{Lee}. It is easy to generalize the above calculations to the general
$(M,N)$ cases, and it can be shown that the topological charge of these field
configuations is $k=M+N$ \cite{Lee}.

\subsection{$SL(2,C)$ CFTW $k$-instantons and jumping lines}

For another subset of $k$-instanton field configurations, one chooses
$\lambda_{i}=\lambda_{i}e_{0}$ (with $\lambda_{i}$ a \textit{complex }number)
and$\ y_{i}$ to be a \textit{biquaternion} in Eq.(\ref{delta}). It is
important to note that for these choices, the constraints in Eq.(\ref{dof})
are still satisfied \textit{without} turning on the off-diagonal elements
$y_{ij}$ in Eq.(\ref{a}). It can be shown that, for these field
configurations, there are non-removable singularities which are zeros ($x\in
J$ ) of%
\begin{equation}
\phi=1+\frac{\lambda_{1}\lambda_{1}^{\circledast}}{|x-y_{1}|_{c}^{2}%
}+...+\frac{\lambda_{k}\lambda_{k}^{\circledast}}{|x-y_{k}|_{c}^{2}},
\label{k}%
\end{equation}
or%
\begin{equation}
\det\Delta(x)^{\circledast}\Delta(x)=|x-y_{1}|_{c}^{2}|x-y_{2}|_{c}^{2}%
\cdot\cdot\cdot|x-y_{k}|_{c}^{2}\phi=P_{2k}(x)+iP_{2k-1}(x)=0. \label{jump}%
\end{equation}
For the $k$-instanton case, one encounters intersections of zeros of
$P_{2k}(x)$ and $P_{2k-1}(x)$ polynomials with degrees $2k$ and $2k-1$
respectively%
\begin{equation}
P_{2k}(x)=0,\text{ \ }P_{2k-1}(x)=0. \label{pp}%
\end{equation}
These new singularities can not be gauged away and do not show up in the field
configurations of $SU(2)$ $k$-instantons. Mathematically, the existence of
singular structures of the non-compact $SL(2,C)$ SDYM field configurations is
consistent with the inclusion of "sheaves" by Frenkel-Jardim \cite{math2}
recently, rather than just the restricted notion of "vector bundles", in the
one to one correspondence between ASDYM and certain algebraic geometric objects.

\section{Acknowledgments}

This talk is based on a collaboration paper with S.H. Lai and I.H. Tsai. I
thank the financial support of MoST, Taiwan.

\end{document}